\documentclass{iaus}
\usepackage{graphicx}
\title[The Oe star HD\,155806]                 
{Is the wind of the Oe star HD\,155806 \\ magnetically confined?}

\author[Fullerton et al.]                  
{  A. W. Fullerton$^1$,
      V. Petit$^{2}$,
      S. Bagnulo$^3$,
      G. A. Wade$^4$,\\
 \and the MiMeS Collaboration
}
\affiliation{$^1$Space Telescope Science Institute,                  
                 3700 San Martin Drive, 
		 Baltimore, MD 21218, USA                            \\
		 email: {\tt fullerton@stsci.edu}                    \\[\affilskip]
             $^2$Dept. of Geology \& Astronomy,
	         West Chester University,                             
	         West Chester, PA 19383, USA                         \\
		 email: {\tt VPetit@wcupa.edu}                       \\[\affilskip] 
	     $^3$Armagh Observatory,  
	         College Hill, Armagh, BT61 9DG, Northern Ireland    \\
	         email: {\tt sba@arm.ac.uk}                          \\[\affilskip]
             $^4$Dept. of Physics,
	         Royal Military College of Canada,
		 Kingston, ON, K7K~4B4, Canada                       \\
		 email: {\tt Gregg.Wade@rmc.ca}                           
	     } 
\pubyear{2010}
\volume{272}                                    
\pagerange{XXX--XXX}
\jname{Active OB stars: structure, evolution, mass loss, and critical limits}
\editors{C. Neiner, G. Wade, G. Meynet \& G. Peters, eds.}
\begin{document}
\maketitle
\begin{abstract}
Spectropolarimetric observations of HD\,155806 -- the hottest Galactic Oe star -- 
were obtained with CFHT/ESPaDOnS to test the hypothesis that disk signatures in its
spectrum are due to magnetic channeling and confinement of its stellar wind.
We did not detect a dipole field of sufficient strength to confine the wind, and
could not confirm previous reports of a magnetic detection.
It appears that stellar magnetism is not responsible for producing the disk of HD\,155806.
\end{abstract}
\vspace*{-0.25 in}
\section{Why Are There Oe Stars?}
Oe stars are a rare subset of the O-type stars that exhibit double-peaked or central 
emission in their Balmer lines. 
This emission-line morphology is distinct from signatures of stellar winds, and is 
conventionally attributed to a circumstellar disk.
Although Oe stars are usually considered to be a continuation of the Be phenomenon 
toward hotter spectral types, it is difficult to understand how stable disks can coexist 
with the increasingly strong stellar winds typical of O-type stars.  

A plausible explanation is that the disk is maintained by a large-scale (e.g., dipolar) 
magnetic field that channels outflowing wind material toward the magnetic equator, in
which case the disk and the wind are really a single entity.
A straightforward test of this hypothesis is to search for dipolar magnetic fields
of sufficient strength to confine the wind.
Since HD\,155806 is the hottest Galactic Oe star currently known
(O7.5~V[n]e according to \cite[Walborn 1973]{Walborn73}), it should have 
the strongest wind; and would therefore require the largest magnetic field 
to confine it.
Moreover, its comparatively narrow photospheric lines and brightness 
enhance the detectability of large-scale magnetic fields.
The hypothesis of magnetic confinement was bolstered when
\cite[Hubrig et al. (2007)]{Hubrig07} 
reported the detection of a magnetic field with longitudinal strength of
$\langle B_z \rangle = -155 \pm 37$ G, though 
\cite[Hubrig et al. (2008)]{Hubrig08} 
did not detect a significant field in subsequent observations.

\section{Observations and Analysis}
\begin{figure}[t]
\begin{center}
 \includegraphics[width=4.0in]{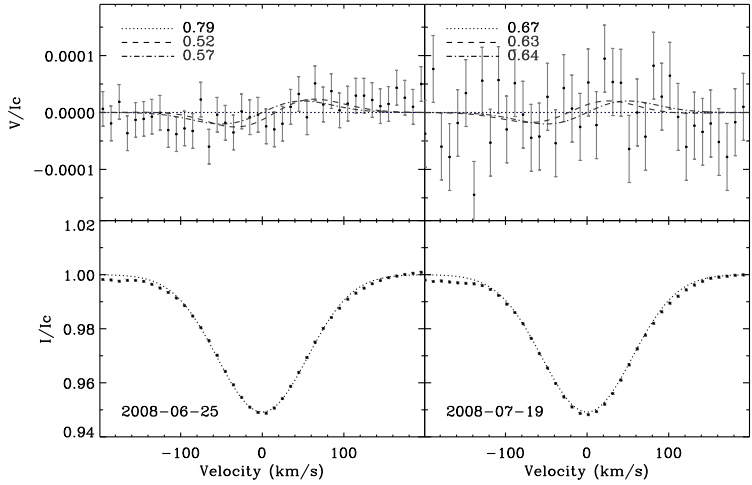}  
 \caption{
   \textit{Lower:}  Mean Stokes I absorption line profiles of HD\,155806 obtained
          with ESPaDOnS.
   \textit{Upper:} Stokes V profiles.  The best fit of a dipole
          model to an individual observation is shown in dashed lines and its corresponding likelihood is
	  labelled, while the best fit to both observations is shown in dot-dashes lines.  
	  Formally, the individual fits give
	  $\langle B_z \rangle = (-55 \pm 19,  8 \pm 37)$~G, but neither is more significant 
	  than the fit for the null hypothesis of ``no field" (dotted line).
	 } 
\label{fig1}
\end{center}
\end{figure}

We used CFHT/ESPaDOnS to obtain high S/N spectropolarimetric observations of 
HD\,155806 on 2008-06-25 and 2008-07-19.
The lower panels of Fig.~\ref{fig1} show the mean absorption line profiles obtained with 
the least-squares deconvolution technique \cite[(Donati et al. 1997)]{Donati97}, 
while the upper panels show the distribution
of circularly polarized light across the mean profiles.
Since deviations as large as the ones observed would occur (73\%, 17\%) of the time
for the observations in (2008 June, 2008 July), these data provide no evidence for the
presence of a magnetic field.
In addition, our reanalysis of the 5 VLT/FORS1 grism observations discussed by 
\cite[Hubrig et al. (2008)]{Hubrig08} resulted in non-detections.
{\bf Thus, we cannot confirm the detection of a significant longitudinal field in
HD\,155806.}
A Bayesian analysis of the ESPaDoNs data shows that a dipolar magnetic field 
cannot be stronger than (199, 425) G with a confidence of (63.3, 95.4)\%.

\section{Implications}
For HD\,155806, a wind magnetic confinement parameter of 
$\eta_\star \equiv { {B^2_{\rm eq}\,R_\star^2 } / { \dot{M}\,v_\infty } } \ge 1 $
(\cite[ud-Doula \& Owocki 2002]{udDoula02})
implies that a dipolar field of {\it at least} 235~G is required to confine the wind.
Since upper limits on the {\it non-detection} of a magnetic field in the ESPaDOnS
spectra are comparable, we conclude that the disk of HD\,155806 is not caused 
by magnetic confinement of its wind.
Similarly, {\cite[Naz\'e, et al. (2010)]{Naze10}} did not detect the hard X-ray 
component in HD\,155806 that should be produced in such a hot star by the collision 
of channelled streams of wind emerging from opposite magnetic hemispheres.
Evidently the disk signatures in the spectrum of HD\,155806 require a different 
explanation.

\end{document}